\documentclass[aps,prd,secnumarabic,amssymb, amsmath,nobibnotes,nofootinbib,twocolumn]{revtex4}
\usepackage{amsfonts,amsmath,hyperref,url, color}
\usepackage{bm, bbm}
\usepackage{graphicx}

\newcommand{\be}{\begin{equation}}
\newcommand{\ee}{\end{equation}}
\newcommand{\bea}{\begin{eqnarray}}
\newcommand{\eea}{\end{eqnarray}}
\newcommand{\bit}{\begin{itemize}}
\newcommand{\eit}{\end{itemize}}

\usepackage{color}

\begin{document}

\title{Accelerated horizons and Planck-scale kinematics}
\author{Michele Arzano}
\email{michele.arzano@roma1.infn.it}
\affiliation{Dipartimento di Fisica and INFN,`Sapienza' University of Rome, P.le A. Moro 2, 00185 Roma, EU }
\author{Matteo Laudonio}
\affiliation{Dipartimento di Fisica `Sapienza' University of Rome, P.le A. Moro 2, 00185 Roma, EU}

\begin{abstract}

We extend the concept of accelerated horizons to the framework of deformed relativistic kinematics at the Planck scale. We show that the non-trivial effects due to symmetry deformation manifest in a finite blueshift for field modes as measured by a Rindler observer approaching the horizon. We investigate whether, at a field theoretic level, this effect could manifest in the possibility of a finite horizon contribution to the entropy, a sort of covariant brick-wall. In the specific model of symmetry deformation considered it will turn out that a non-diverging density of modes close to the horizon can be achieved only by introducing a momentum space measure which violates Lorentz invariance.

\end{abstract}

\maketitle

\section{Introduction}

One of the outstanding puzzles in the search for a theory of quantum gravity concerns the fate of information when a black hole undergoes quantum radiance \cite{Mathur:2008wi}. Loosely speaking the information on matter and radiation fallen behind the black hole horizon contributes to the entropy budget of the black hole according to the generalized second law of thermodynamics \cite{Bekenstein:1974ax}. It was indeed an argument relying on the information carried by a particle falling into a black hole and quantum uncertainty which led Bekenstein to propose the celebrated relationship between horizon area and entropy \cite{Bekenstein:1973ur}. The Bekenstein-Hawking entropy-area law
\be\label{BES}
S = \frac{A}{4 L^2_p}\,,
\ee
with $L_p=\sqrt{\frac{\hbar G}{c^3}}\sim 10^{-35}m$ the Planck length, has been derived to date in various approaches to quantum gravity (see \cite{Carlip:2008rk} for a comprehensive list) but a definite understanding of the nature of the entropy in \eqref{BES} remains elusive.

An illuminating derivation of the entropy-area law as thermodynamic entropy of a field in the presence of a black hole horizon is due to 't Hooft \cite{tHooft:1985vk}. This result relies on field mode counting which leads to a divergent density of states at the horizon. In order to have a finite result a regulator in terms of Dirichlet boundary conditions on the field at finite distance from the horizon has to be imposed, effectively creating a ``brick wall". The regulated entropy scales as the {\it area} of the wall, unlike the entropy of a field in Minkowski space which is extensive and scales as the volume of the box in which the field is contained. One can then tune the distance of the wall from the horizon and reproduce exactly the Bekenstein-Hawking relation \eqref{BES}. A common view is that such brick wall regularization is a mathematical tool which models unknown Planckian physics features near the horizon which effectively provide a UV cut-off for the modes of the field. Since close enough to the horizon the space looks locally flat, to a static observer in this limit the black hole horizon will become indistinguishable from a Rindler horizon. Indeed a calculation of the number of modes of a field near a Rindler horizon \cite{Susskind-Uglum:1994, Brustein:2010ms} shows the same type of divergence. It is thus reasonable to expect that the same type of Planckian physics which renders the brick-wall entropy finite should also lead to a finite entropy density for a field, {\it in flat space}, in the presence of a Rindler horizon.

The existence of an entropy associated to the Rindler horizons is also the key ingredient in Jacobson's derivation of Einstein's equation as an equation of state \cite{Jacobson:1995ab}. This suggestive result, besides a {\it finite} entropy density associated to the horizon, relies on the assumption of a local Unruh temperature $T_U$ and the validity of an entropy balance relation $\delta S = \delta E/ T_U$ where $\delta E$ is the energy flow across the horizon. Whether the entropy density one associates with the Rindler horizon is due to quantum correlations or it is thermodynamic in nature its value in ordinary quantum field theory is divergent. In agreement with the picture discussed above, it is expected that the regulator which renders finite the horizon entropy density is provided by quantum properties of space-time at the Planck scale (see e.g. \cite{Chirco:2014saa, Arzano:2017mdp,Arzano:2017goa}). 

In this work we explore the possibility that Planck scale effects, modelled by a Hopf algebra deformation of the symmetries of Rindler space, could naturally provide the needed mode cut-off at the horizon and, eventually, lead to a finite entropy density associated to a Rindler horizon. In the next Section we start by characterizing Rindler modes and horizon in terms the action of symmetries generated by the Weyl-Poincar\'e algebra. This will prepare the ground for a derivation of the same concepts in the context of a quantum deformation of the Weyl-Poincar\'e algebra. In Section 3 we recall how mode counting in Rindler space leads to an area law for the thermodynamic entropy of a field once the divergent contribution from the horizon is regulated by a ``brick wall".  In Section 4 we introduce a deformation of the Weyl-Poincar\'e algebra based on a mathematical procedure know as ``twist". From the deformed commutators between translation, boosts and dilation generators we derive finite dilation and boosts transformations for four-momenta. We will use these transformations to define deformed Rindler translation generators and the blueshift of Rindler modes near the horizon, obtaining a Planckian {\it upper bound} to the frequency of these modes. In Section 5 we perform field mode counting in the deformed Rindler setting. We show that the density of states of the deformed field depends dramatically on the choice of integration measure in momentum space. Using an ordinary measure, and thus breaking the deformed Lorentz symmetry, we obtain a {\it finite} density of states while adopting a covariant, deformed, measure leads to a divergent density of states as in ordinary Rindler space. The last Section is devoted to a discussion of the results presented and to a brief  outlook of further developments.

\section{Accelerated horizons and the Weyl-Poincar\'e group}
Static observers in Minkowski space are observers whose four-velocities are identical to the time-translation Killing vector. A photon emitted with frequency $\omega$ by a static observer will be detected by another static observer with the same frequency. An inertial observer boosted in a given direction by a parameter $\phi$ will measure the frequency of the same photon redshifted by a factor $e^{-\phi}$: $\omega' = e^{-\phi} \omega$. Since $\cosh \phi=\gamma=\frac{1}{\sqrt{1-\beta^2}}$ and $\sinh \phi = \gamma\beta$  this statement amounts to the familiar redshift formula for the frequency of a photon as measured by an observer moving with velocity $\beta$
\be
\omega' = \sqrt{\frac{1-\beta}{1+\beta}} \omega\,.
\ee
The fact that the time flow along an accelerated trajectory in Minkowski space can be parametrized by the rapidity of a boost transformation is well known. An observer moving along such trajectory will see a photon emitted by a static Minkowski observer continuously redshifted by a factor $e^{- \alpha \tau}$ where $\tau$ is her proper time and $\alpha$ a constant with dimensions of inverse length in natural units. Accordingly the four velocity along such trajectory can be written as an inverse\footnote{Recall that if $p_{\mu}$ is the four-momentum measured by a given observer $p_{\mu}U^{\mu}$ is a conserved quantity and thus if $p_{\mu}$ is boosted $U^{\mu}$ is anti-boosted.} boost of the static four-velocity $U^{\mu}=(1,0,0,0)$ by the same dimensionless parameter $\alpha\tau$

\be\label{4v}
U^\mu=(\cosh \alpha\tau,\sinh \alpha\tau,0,0) \,.
\ee
Taking the proper time derivative of such four velocity one can immediately see that $\alpha$ is the magnitude of the four-acceleration of the observer.

Focusing for simplicity on the case of $1+1$ dimensions, we see, integrating \eqref{4v}, that the worldline of the accelerated observer can be parametrized in Minkowski coordinates by $t(\tau)$ and $x(\tau)$ obeying the relation
\be
\label{accelerated worldline}
-t(\tau)^2+x(\tau)^2 =\frac{1}{\alpha^2}\,, 
\ee
describing a two-dimensional hyperbola: the boost orbit of the vector $(0,1/\alpha)$. It is evident that a rescaling of the space-time coordinates $t(\tau)$ and $x(\tau)$ will move us to another trajectory corresponding to a {\it different} acceleration. Such rescaling can be obtained by acting with a {\it dilation} transformation $(t,x)\rightarrow (t',x')= e^{\delta} (t,x)$ generated by the operator $D= -i x^{\mu} \partial_{\mu}$. Geometrically the effect of such dilation is that of moving from a boost orbit intersecting the $x$-axis in $1/\alpha$ to another intersecting it at $e^{\delta}/\alpha$. The dilation parameter can be used to define a spatial coordinate on the Rindler wedge if we rewrite it in terms of an acceleration $a$ as $\delta= a \xi$ so that now 
\bea
t & = &\frac{1}{a}e^{a\xi}\sinh{a\eta}\\
x & = &\frac{1}{a}e^{a\xi}\cosh{a\eta} 
\eea
and for a given $\xi$ the acceleration along the trajectory parametrized by $\eta$ is $\alpha = a\, e^{-a\xi}$. We see that for $\xi\rightarrow -\infty$ the acceleration diverges indicating that the light-cone acts as a horizon for Rindler observers.

A crucial feature we would like to stress is that transformations generated by the Weyl-Poincar\'e group (the Poincar\'e group extended to include a dilation transformation, in our case generated by $D= -i ( t \partial_t + \, x \partial_x )$, provide enough structure to describe Rindler observers, redshift and horizons. 
To do so let us recall that the Weyl-Poincar\'e algebra in 1+1 dimensions is given by the following commutators
\begin{align}
\bigl[P_t,P_x\bigr]& =0 \,,\,\,\, \bigl[D,N\bigr] =0\\
\bigl[N,P_t\bigr]& =iP_x  \,,\,\,\, \bigl[N, P_x\bigr]  =iP_t\\
\bigl[D,P_t\bigr] & =iP_t\,,\,\,\, \bigl[D,P_x\bigr]  =iP_x \,.
\end{align}
Notice how this algebra contains {\it two abelian subalgebras} spanned by $\{P_t,P_x\}$ and $\{D,N\}$. Thus, besides the usual representation in terms of derivatives with respect to space and time coordinates $P_{t,x}=-i \partial_{t,x}$, we have an alternative one in terms of Rindler space and time coordinates $\xi$ and $\eta$ where Rindler space and time translation generators $P_\xi = a D $ and $P_\eta = a N$ are given by derivatives with respect to $\xi$ and $\eta$, respectively:
\begin{align}
P_\xi & = -i \partial_\xi\\
P_\eta & = -i \partial_\eta\\
P_t & = -i e^{-a\xi}(\cosh{a\eta}\,\partial_\eta -\sinh{a\eta}\,\partial_\xi)\\
P_x & = -i e^{-a\xi}(-\sinh{a\eta}\,\partial_\eta+\cosh{a\eta}\,\partial_\xi)\,. 
\end{align}
Inverting the last two relation, we have
\begin{align}\label{RindMod1}
P_\eta & =  e^{a\xi}(\cosh{a\eta}\,P_t + \sinh{a\eta}\,P_x)\\
P_\xi & = e^{a\xi}(\sinh{a\eta}\,P_t+\cosh{a\eta}\,P_x)\,, 
\end{align}
showing that Rindler translation generators (and consequently the associated modes) can be obtained from the action of the ``Weyl-Lorentz" subgroup of the Weyl-Poincar\'e group comprised of dilations and boosts. From the maps above one can immediately obtain the Fourier transform of the Rindler wave operator simply expressing the usual Casimir of the Poincar\'e algebra in terms of Rindler translation generators
\be
\mathcal{C} = - P^2_0 + P^2_x = e^{-2a \xi} (- P^2_\eta + P^2_\xi)\,, 
\ee
which in terms of eigenvalues of $P_\eta$ and $P_\xi$ translates in the dispersion relation for Rindler frequency and wave-number. Notice that the Casimir operator $\mathcal{C}$ {\it is not} an invariant of the Poincar\'e-Weyl group and indeed dilation transformations change the mass eigenvalue of $\mathcal{C}$ thus allowing to jump from one mass-shell to the other. However, as it is well known, mass-shells of massless particles are invariant under the full Poincar\'e-Weyl group.

Static Rindler observers are characterized by a four-velocity proportional to the four-vector $P^{\mu}_\eta = e^{a\xi}(\cosh{a\eta}\,, \sinh{a\eta})$. Such four-velocity can be obtained by an inverse boost by $\alpha\tau$ of the four-momentum of a Minkowski observer at rest $(m,0)$ divided by its mass:
\be
U^{\mu} = (\cosh{\alpha\tau}\,, \sinh{\alpha\tau})
\ee
The proper time of an observer with acceleration $\alpha$ will be given by $\tau = \frac{a}{\alpha}\eta$ and her trajectory will be characterized by a fixed $\xi = \frac{1}{a} \log\frac{a}{\alpha}$ as shown in our discussion above. Thus, for the four-velocity of such an accelerated observer, we will have
\be
U_{\xi}^{\mu} =  e^{- a\xi} P^{\mu}_\eta\,.
\ee
The conserved energy of a photon with four-momentum $k_{\mu}$ with respect to the Rindler time is $E_{\eta} = k_{\mu} P^{\mu}_\eta$ while its frequency measured by the observer with acceleration $\alpha = a\, e^{-a\xi}$ is given by $\omega_{\xi} =  k_{\mu} U_{\xi}^{\mu} = e^{- a\xi} E_{\eta}$ so that
\be\label{RshiftO}
\omega_{\xi} = e^{- a\xi} E_{\eta}\,.
\ee
The photon's frequency will appear infinitely blueshifted when measured at $\xi=-\infty$ in analogy with the frequency shift of photons for static observers approaching the event horizon of a Schwarzschild black hole. This reflects the fact that the region of the Rindler wedge at $\xi=-\infty$ is a casual horizon for accelerated observers.

What is more interesting for our purposes is that the same dilation transformation which allows us to define Rindler translation generators from Minkowski ones is responsible for the infinite Doppler-shift which ultimately signals the presence of the horizon. Intuitively this is easily understood by considering the horizon as the locus of observers with infinite acceleration. As discussed above, dilation transformations change the value of acceleration from zero at $\xi=\infty$, reached by an infinite dilation, to infinity at the horizon, $\xi=-\infty$, corresponding to an inverse dilatation by an infinite parameter $\xi$. In what follows we will characterize the Rindler Doppler shift in terms of dilation transformations and the accelerated horizon in terms of an {\it infinite inverse dilation}. 

\section{Area-law from flat space-time}

In $3+1$-dimensional Minkowski space the number of modes of a massless scalar field in a box of size $L$ is calculated starting from imposing the presence of nodes on the walls of the box, i.e.
\be
\label{number of nodes Minkowski 3+1}
k_i =N_i\frac{\pi}{L} \quad \text{where}\quad i=1,2,3 \,\,,\quad N_i=1,2,3,\ldots
\ee
The infinitesimal volume in the space of wave numbers is $dV_k=\frac{4\pi k^2 dk}{8}$, where we have restricted to the positive octant.  The infinitesimal density of states is given by $dV_k$ divided by the constant spacing of modes $\left(\frac{\pi}{L}\right)^3$:
\be
\label{infinitesimal density of states Minkowski}
dn= \frac{L^3}{2\pi^2}k^2 dk = \frac{L^3}{2\pi^2}E^2 dE,
\ee
where we used  the trivial dispersion relation for a massless field $E=|\vec{k}|=k$. The density of states is given by integrating \eqref{infinitesimal density of states Minkowski} over the energy
\be
\label{density of states Minkowski 3+1}
n(E)= \frac{L^3E^3}{6 \pi^2}.
\ee
From the partition function associated to this density of states one obtains the well known {\it extensive} behaviour of the thermodynamic entropy of the field which scales as the volume of the box $L^3$. From an holographic perspective \cite{Bousso:2002ju} this is a sign that local quantum field theory possesses too many degrees of freedom and that some kind of truncation \cite{Yurtsever:2003ii} dictated by quantum gravity effects might produce an entropy scaling as the {\it area} of the box, in accordance with various versions of the {\it holographic entropy bound} \cite{Bekenstein:1980jp,Bousso:1999xy}.

It should be noted however that, even in the absence of gravity, local quantum field theory exhibits a type of entropy-area relation, although this relies on the introduction of a UV regulator. This happens in Rindler space and the derivation \cite{Susskind-Uglum:1994} is essentially an adaptation of 't Hooft brick wall calculation \cite{tHooft:1985vk}  to the case of infinite black hole mass. Here we show how this result can be obtained already with a simple counting argument similar to the one given above. 

Let us consider first the case of 1+1 dimensional Rindler space.  With a constant wavelength $\lambda$ we know that the number of angular cycles over a space interval $\Delta x$ is given by
\be
\Delta \phi = \frac{2\pi \Delta x}{\lambda}\,.
\ee
If the wavelength changes over space we can define an infinitesimal version of the above equation. In particular for Rindler space the wavenumber is given by $k(\xi) = e^{-a\xi} k$ where $k$ is the ``comoving" wavenumber (conserved charge associated to space translations). The infinitesimal number of angular cycles is thus given by
\be
d \phi = \frac{2\pi d \xi}{\lambda(\xi)} = k(\xi) d \xi\,,
\ee
from which
\be
\phi = \int^L_{\xi_{\mathrm{min}}} k(\xi) \sqrt{g_{\xi\xi}}\, d \xi=\int^L_{\xi_{\mathrm{min}}} k\, d \xi\,,
\ee
where $\xi_{\mathrm{min}}$ is a regulator needed to avoid the divergence at $\xi = -\infty$. This leads to the standing wave condition 
\be\label{rswc}
 \int^L_{\xi_{\mathrm{min}}} k  d \xi = N \pi\,.
\ee
The infinitesimal volume in Rindler momentum space is given by
\be
d V^R_k(\xi) =  e^{-a \xi} dk\,,
\ee
and, in order to derive the number of modes within such infinitesimal volume, we have to divide it by the volume of each mode. We want to find a generalization of the volume $\pi/L = dk/dN$ occupied by each mode in the Minkowski case. In our case we note that 
\be
\frac{dN}{d k(\xi)} = \frac{dN}{d\xi} \frac{d \xi}{d k(\xi)}\,,  
\ee
since from \eqref{rswc} $\frac{dN}{d \xi} = \frac{k}{\pi}$ we have that
\be
\label{N/k xi dir}
\frac{dN}{d k(\xi)} = \frac{k}{\pi} \frac{d \xi}{d k(\xi)}\,.  
\ee
The infinitesimal number of modes will be then given by
\be
dn= d V^R_k(\xi) \frac{dN}{d k(\xi)} =  \frac{k}{\pi}\, d\xi\,.
\ee
Generalizing to the $3+1$-dimensional case the number of nodes in the $\xi$-direction is still given by eq. \eqref{rswc}, while the number of nodes in the orthogonal plane $(y,z)$ is constant as in Minkowski space:
\be
k_\perp=k_i =N_i \frac{\pi}{L}  \quad i=y,z\,,\,\, N_i=1,2,3,\ldots\,. 
\ee
Now, the infinitesimal volume in Rindler momentum space is given by
\be
\label{volume Rindler space}
dV_k^R=\frac{e^{-a\xi}\, 2\pi k_\perp \, dk_\perp \, dk}{4} \,,
\ee
and thus the density of states is obtained from \eqref{volume Rindler space} divided by the volume occupied by each mode in $\xi$-direction \eqref{N/k xi dir}
and divided by the volume occupied by orthogonal modes $\frac{L^2}{\pi^2}$
\be
\label{infinitesimal density of states Rindler off-shell}
dn = \, \frac{L^2 \, k_\perp \, k}{2\pi^2} \, dk_\perp d\xi \, .
\ee
Introducing the Rindler dispersion relation $E^2=k^2+e^{2a\xi}k_\perp^2$, and integrating over $k_\perp$ we find
\begin{align}
d n
& =\, \frac{L^2}{2\pi^2} \, d\xi \, \int_0^{E e^{-a\xi}} dk_\perp \, k_\perp \, \sqrt{E^2-k_\perp^2 e^{2a\xi}}= \notag \\
\label{infinitesimal density of states Rindler}
& = \, \frac{E^3 \, L^2}{6\pi^2} \, e^{-2a\xi} \, d\xi \,.
\end{align}
To obtain the expression for the number of states as in the Minkowski case we should integrate over the spatial coordinate $\xi$. We immediately see that \eqref{infinitesimal density of states Rindler} will give an infinite result and two types of regularization are needed. One is the usual IR regularization which puts the system in a box for large $\xi$, as in Minkowski space. Unlike the Minkowski case however we have now a new divergent contribution coming from the modes {\it near the horizon} i.e. when $\xi\rightarrow -\infty$, and thus we have to introduce a ``brick wall" regulator so that the field has nodes at $\xi_{min}$ close to the horizon. Thus to get the number of states we integrate \eqref{infinitesimal density of states Rindler} assuming the presence of a brick wall i.e. the restriction $\xi > \xi_{min}$ and a large box at  $x_{max} = R$ so that $\xi<\frac{\log aR}{a}$; the density of states is then given by
\begin{align}
n(E)
& =\, \frac{E^3 \, L^2}{6\pi^2} \int_{\xi_{min}}^{\log (aR)/a} d\xi \, e^{-2a\xi}\notag \\
\label{density of states Rindler 3+1}
& =\, \frac{E^3 \, L^2}{12\pi^2}\frac{1}{a}\left[ e^{-2a\xi_{min}}-\frac{1}{(aR)^2} \right]\,.
\end{align}
We can now calculate the entropy associated to this density of states. Starting from the logarithm of the thermodynamical partition function
\be
\log Q=\beta\int_0^\infty \, dE \frac{n(E)}{e^{\beta E}-1}.
\ee
the entropy is derived from the usual relation
\be
S = -  \beta^2\frac{\partial }{\partial \beta}\frac{\log Q}{\beta}\,,
\ee
and, using the density of states \eqref{density of states Rindler 3+1}, one obtaines
\be
S=\frac{\pi^2}{45}\frac{L^2}{a\beta^3}\left[ e^{-2a\xi_{min}}-\frac{1}{(aR)^2} \right]\,.
\ee
We see that the UV contribution due to brick wall near the Rindler horizon
\be
S_{wall}=\frac{\pi^2}{45}\frac{L^2}{a\beta^3}e^{-2a\xi_{min}}
\ee
scales as an area and the value $\xi_{min}$ can be adjusted so that, at the Hawking temperature, one has an entropy density $\sigma_{wall} = S_{wall}/L^2=1/4 L^2_p$ exactly reproducing the entropy density associated to the Bekenstein-Hawking area law. 

For our purposes it will be useful to recall that the above results on state counting for a massless field can be recast in a covariant version by considering the state space of a massless relativistic point particle. In this case the infinitesimal number of states is given 
\be
dn = \frac{1}{(2\pi)^3} 2E\, dt\, d^3x\, \delta(t)\, d^4p\, \delta(\mathcal{C})\, \theta(E)
\ee
where $ 2E\, dt\, d^3 x\, \delta(t)$ is the covariant measure in configuration space and $d^4p\, \delta(\mathcal{C})\, \theta(E)$ is the covariant measure on the physical momentum space, i.e. the positive energy mass-shell determined by the Casimir relation $\mathcal{C}$.
In Minkowski space, integrating over a spatial volume $V$ and up to an energy $E$, we have
\be
n(E)=\frac{V}{(2\pi)^3}\int_E d^4p \, 2 p_0\, \delta(\mathcal{C})\,\theta(p_0)\,,
\ee
which, carrying out the integration over momentum space, reproduces \eqref{density of states Minkowski 3+1}. In Rindler space the integral over momentum space becomes
\be
 \int \, dp_\eta dp_\xi dp^2_\perp \, 2p_\eta e^{-3a\xi} \,  \delta(\mathcal{C})\theta(p_\eta) 
\ee
with
\be
\label{cas.Rind.}
\mathcal{C}=- e^{-2a\xi}p_\eta^2 + e^{-2a\xi}p_\xi^2 + p_\perp^2.
\ee
so that 
\be\label{nRindU}
n(E) = \frac{V_\perp}{(2\pi)^3}\int_{\mathbb{R}} \, d\xi \int \, dp_\eta dp_\xi dp^2_\perp \, 2p_\eta e^{-2a\xi} \,  \delta(\mathcal{C})\theta(p_\eta)\,.
\ee
This integral can be evaluated and leads to the known divergent result 
\be\label{RindModM}
n(E)=\frac{V_\perp}{(2\pi)^3}\frac{4\pi}{3}E^3\int_{-\infty}^\infty d\xi \, e^{-2a\xi}\,,
\ee
which, appropriately, regularized gives \eqref{density of states Rindler 3+1}. 

In the following Sections we will introduce deformations of relativistic symmetries characterized by a fundamental, observer independent, length scale which affects the way boosts (and dilations) act on momentum space. These deformations profoundly alter the structure of momentum space introducing in a subtle way a UV cut-off. We will generalize the discussions on state counting above in order to see whether the presence of such fundamental scale can provide a natural cut-off in \eqref{RindModM} and possibly a finite entropy density associated to a Rindler horizon.

\section{Deformed symmetries: the twisted Weyl-Poincar\'e algebra}

As mentioned in the Introduction, it is a widely accepted view that the divergent entropy density we just derived for Rindler space should be somewhat naturally regularized when quantum properties of space-time are taken into account \cite{Chirco:2014saa,Arzano:2017mdp,Arzano:2017goa}. It is thus interesting to look at possible new features of accelerated horizons which might be determined by Planck scale physics. Deformations of relativistic symmetries using quantum group techniques \cite{Lukierski:1991pn, Lukierski:1992dt, Majid:1994cy} have been widely studied in the past years as possible candidates for a ``flat space-time limit" of quantum gravity \cite{KowalskiGlikman:2004qa}. The role of these deformation is particularly evident for quantum gravity in $2+1$-dimensions \cite{Bais:2002ye, Freidel:2003sp, Arzano:2013sta, Arzano:2014ppa} and it has been speculated that they might be relevant also in the more realistic $3+1$ dimensional case \cite{AmelinoCamelia:2003xp}.

The vast majority of studies in the literature have so far focused on the deformations of the Poincar\'e algebra \cite{Lukierski:1991pn,Lukierski:1992dt,Lukierski:1993wxa,Majid:1994cy,Arzano:2007ef, Arzano:2010jw, Kowalski-Glikman:2017ifs}. As we have seen in Section 2, in order to describe the kinematics of accelerated observers one has to include dilations in the picture and thus switch the attention to deformations of the Weyl-Poincar\'e algebra\footnote{For earlier attempts to the study of Rindler space and accelerated observers starting from  $\kappa$-Minkowski space and the associated deformed $\kappa$-Poincar\'e algebra see  \cite{Kim:2007pu,KowalskiGlikman:2009ff, Harikumar:2012yu}.}. These have been recently considered in the literature \cite{Aschieri:2017ost} in the context of the so-called twist deformations (see also \cite{Bu:2006dm,Juric:2012xt}). Quite interestingly it turns out that the type of deformation of the Weyl-Poincar\'e algebra derived in \cite{Aschieri:2017ost} reproduces, in the Poincar\'e sub-algebra sector, the features of a deformation of the Poincar\'e algebra obtained in \cite{Magueijo:2001cr} in the early days of the studies of ``doubly special relativity" (DSR) theories\footnote{Another deformation of the Weyl-Poincar\'e algebra, different from the one we consider in this work, was proposed in \cite{Ballesteros:2003kz} in light of possible applications as a DSR model. In \cite{Ballesteros:2003kz} finite boosts transformations were derived but finite dilations transformations were not analyzed.} \cite{AmelinoCamelia:2000ge, AmelinoCamelia:2000mn}.

Let us start by recalling the commutators of the ordinary Weyl-Poincar\'e algebra:
\begin{align}
\label{PW.comm.1}
& [P_\mu,P_\nu]=0 \,\,,\quad [P_\mu,M_{\rho\nu}]=i\bigl(\eta_{\mu\rho}P_\nu-\eta_{\mu\nu}P_\rho \bigr) \\
&\bigl[M_{\mu\nu},M_{\rho\sigma}\bigr] =i\bigl(\eta_{\mu\sigma}M_{\nu\rho}-\eta_{\mu\sigma}M_{\nu\rho}+\eta_{\nu\rho}M_{\mu\rho}-\eta_{\nu\rho}M_{\mu\sigma}\bigr) \\
\label{PW.comm.2}
&[D,P_\mu]=iP_\mu \,\,, \quad \bigl[D,M_{\mu\nu}\bigr]=0\,.
\end{align}
Denoting with $W$ a generic generator of the algebra $\eqref{PW.comm.1}$, the twist procedure consists in defining new generators 
\be
W^{\mathcal{F}} = \bar{f^\alpha}(W)\bar{f_\alpha}
\ee
where the twist element $\mathcal{F}=f^\alpha\otimes f_\alpha$ is given by
\be
\label{Jordanian.twist}
\mathcal{F}=\exp(-i D\otimes \sigma) \,,\quad \sigma=\log\left( 1+\ell P_0 \right).
\ee
and its inverse is denoted by $\mathcal{F}^{-1}=\bar{f^\alpha}\otimes \bar{f_\alpha}$ (see \cite{Aschieri:2017ost} for details). It turns out that such twisting procedure only affects the translation generators and indeed the final result is
\be\label{twistedPWG}
P_\mu^\mathcal{F}=\frac{P_\mu}{1+\ell P_0} \,,\quad M_{\mu\nu}^\mathcal{F}=M_{\mu\nu} \,,\quad D^\mathcal{F}=D.
\ee
It should be noted that an analogous non-linear redefinition of momenta was first proposed by Magueijo and Smolin in \cite{Magueijo:2001cr} in order to derive a deformation of the Poincar\'e algebra in the context of DSR theories. We will return to this connection at the end of this Section.

The feature of the twisted kinematics which will be most relevant for us will be the deformation of the commutators between translation, boosts and dilation generators. Indeed it will be from such relations which we will derive a non-trivial action of finite boosts and dilations which will allow us to define deformed Rindler momenta. In \cite{Aschieri:2017ost} it was observed that in terms of the ``braided" commutator
\be
\label{twist.comm.}
[W^\mathcal{F},V^\mathcal{F}]_\mathcal{F}=W^\mathcal{F}V^\mathcal{F}-(\bar{R}^\alpha(V))^\mathcal{F}(\bar{R}_\alpha(W))^\mathcal{F}\,,
\ee
where $\mathcal{R}^{-1}=\bar{R}^\alpha\otimes \bar{R}_\alpha$ is the inverse if the universal $R$-matrix associated to the twist $\mathcal{F}$, the twisted generators \eqref{twistedPWG} obey an undeformed Weyl-Poincar\'e algebra. This translates in the  {\it deformed} commutators
\begin{align}
\label{def.boost.comm.}
& [M_{\mu\nu}^\mathcal{F},P_\rho^\mathcal{F}]=i\bigl(\eta_{\rho\nu}P_\mu^\mathcal{F} -\eta_{\rho\mu}P_\nu^\mathcal{F}\bigr)-i\ell \delta_{\mu 0}\delta_{\nu i}P_\rho^\mathcal{F}P_i^\mathcal{F} \\
\label{def.dil.comm.}
& [D^\mathcal{F},P_\mu^\mathcal{F}]=iP_\mu^\mathcal{F}-i\ell P_\mu^\mathcal{F}P_0^\mathcal{F}
\end{align}
while all other commutators remain undeformed\footnote{It should be noted that, at a Hopf algebraic level, a crucial role is played by the notions of {\it coproduct} and {\it antipode} which describe, respectively, the action of symmetry generators on tensor product and on dual representations (see \cite{Fuchs:1997jv} for a gentle introduction to Hopf algebras and \cite{Arzano:2014cya} for the role of antipode and coproduct in relativistic quantum theory). These structures affect the multiparticle sector of the theory and, at the kinematical level, are responsible, among other things, for a non-abelian composition rule of four-momenta. For the purposes of the present work this aspect of symmetry deformation will not be relevant and we refer the reader to \cite{Aschieri:2017ost} for more details.}.  An important object, for what concerns relativistic kinematics, is the Casimir invariant of the Poincar\'e subalgebra. It can be immediately seen that the undeformed Casimir $\mathcal{C}=P_\mu P^\mu$ in terms of the twisted translation generators $P_\mu^\mathcal{F}$ becomes
\be
\label{def.cas.}
\mathcal{C}^\mathcal{F}=\frac{(P_\mu P^\mu)^\mathcal{F}}{(1-\ell P_0^\mathcal{F})^2}.
\ee
Using \eqref{def.boost.comm.} and \eqref{def.dil.comm.} we can now derive derive the expressions for finite boosts and dilations. We first boost the four-momentum $(\omega,\vec{k})$ in the $1$-direction by a parameter $\phi$. From \eqref{def.boost.comm.} we have 
\begin{align}
& \frac{d\omega}{d\phi}=-i[N_1,\omega]=k_1(1-\ell \omega) \\
\label{twist.diff.ew.k1}
& \frac{d k_1}{d\phi}=-i[N_1,k_1]=(\omega-\ell k_1 k^1) \\
& \frac{d k_i}{d\phi}=-i[N_1,k_i]=\ell k_1 k_i\,,\,\,\,\,i=2,3 \label{twist.diff.ew.k3}\,.
\end{align}
Substituting the first equation in the second one, we find a second order differential equation for $\omega(\phi)$:
\be
\frac{\partial ^2 \omega}{\partial \phi^2}+2\ell\left( \frac{\partial \omega}{\partial \phi} \right)^2 \frac{1}{(1-\ell\omega)}-\omega(1-\ell\omega)=0.
\ee
By solving the above equation and introducing the solution in eq. \eqref{twist.diff.ew.k1}-\eqref{twist.diff.ew.k3}, we can find solutions with initial conditions $(\omega^0,\vec{k}^0)$ for $\phi=0$:
\begin{align}\label{energy.boost.}
& \omega(\phi)=\frac{\omega^0 \cosh\phi+k_1^0\sinh\phi}{A} \\
& k_1(\phi)=\frac{\omega^0 \sinh\phi+k_1^0\cosh\phi}{A}\label{moment1.boost.}\\
& k_i(\phi)=\frac{k_i^0}{A}\,,\,\,\, i=2,3 \label{moment2.boost.}\,,
\end{align}
where
\be
A=1-\ell\omega^0+\ell\omega^0 \cosh\phi+\ell k_1^0\sinh\phi.
\ee
In the limit $\ell\rightarrow 0$ one recovers the usual form of the Lorentz boosts. The salient feature of these deformed boosts however is that letting the boost parameter going to infinity, $\omega$ and $k_1$ reach a {\it finite limit} at $1/\ell$ while $k_2$ and $k_3$ contract to zero. This is a typical DSR feature and it is analogous to the behaviour of finite boosts of the $\kappa$-Poincar\'e algebra found in \cite{Bruno:2001mw}. Indeed as we mentioned above a similar deformation of the translation sector of the Poincar\'e algebra was considered in \cite{Magueijo:2001cr} where deformed boost transformations identical to  \eqref{energy.boost.} were found.

Unlike the framework of the $\kappa$-Poincar\'e algebra we now have also the tools for deriving a deformed action of dilations on energy and momenta. From \eqref{def.dil.comm.} we have for an infinitesimal dilation of parameter $\delta$:
\begin{align}
& \frac{d \omega}{d\delta}=-i[D,\omega]=\omega(1-\ell \omega) \\
\label{twist.dil.diff.eq}
& \frac{d k_i}{d \delta}=-i[D,k_i]=k_i(1-\ell\omega)\,, \,\,\,\, i=1,2,3\,.
\end{align}
These differential equations can be easily solved for initial values $(\omega^0,\vec{k}^0)$ at $\delta=0$ to get 
\begin{align}
\label{en.mom.dil.}
&\omega(\delta)=\frac{\omega^0}{\omega^0\ell+(1-\omega^0\ell)e^{-\delta}} \\
& k_i(\delta)=\frac{k_i^0}{\omega^0\ell+(1-\omega^0\ell)e^{-\delta}}\,, \,\,\,\, i=1,2,3\,.
\end{align}

\newpage

Notice that in the case of ordinary dilations both energy and momentum diverge for $\delta\rightarrow\infty$ while for $\delta\rightarrow -\infty$ they vanish. In the presence of deformation we witness a saturation effect similar to the one encountered for the twisted boosts. Indeed, while for $\delta \rightarrow -\infty$ both energy and momentum vanish, for $\delta \rightarrow \infty$ we have
\be
 \lim_{\delta\to\infty}\omega(\delta)=\frac{1}{\ell} \qquad  \lim_{\delta\to\infty}k_i(\delta)=\frac{k_i^0}{\ell\omega^0} 
\ee
and in particular for a massless field {\it both energy and spatial momentum saturate at the value of $1/\ell$}.

Before discussing the relevance of this result for a potential generalization for the notion of horizon let us define Rindler translation generators. Our strategy relies on the symmetry arguments given in Section 2. Indeed we assume that Rindler translation generators associated to accelerated observers in the 1-direction are obtained from the deformed Minkowski generators $P^\mathcal{F}_{\mu}$ acting with a deformed boost  in the 1-direction and a dilation in the $\eta-\xi$-plane. This realizes the following $3+1$-dimensional generalization of \eqref{RindMod1} 
\begin{align}
\label{def.Rind.coord.1}
& P^{\mathcal{F}}_\eta=\frac{P^{\mathcal{F}}_0 \cosh a \eta+ P^{\mathcal{F}}_1\sinh a \eta}{\ell P^{\mathcal{F}}_0 \cosh a \eta+\ell  P^{\mathcal{F}}_1\sinh a \eta+(1- P^{\mathcal{F}}_0\ell)e^{- a \xi}} \\
& P^{\mathcal{F}}_\xi=\frac{ P^{\mathcal{F}}_0 \sinh a \eta+ P^{\mathcal{F}}_1\cosh a \eta}{\ell P^{\mathcal{F}}_0 \cosh a \eta+\ell  P^{\mathcal{F}}_1\sinh a \eta+(1- P^{\mathcal{F}}_0\ell)e^{- a \xi}} \\
\label{def.Rind.coord.2}
& P^{\mathcal{F}}_i=\frac{P^{\mathcal{F}}_i}{1-\ell P^{\mathcal{F}}_0+\ell P^{\mathcal{F}}_0 \cosh a \eta+\ell  P^{\mathcal{F}}_1\sinh a \eta}\,,\,\, i=2,3.
\end{align}
where we used numeric indices for $\mu=0,..,3$ for deformed Minkowski generators while we used mixed indices $\eta,\xi$ and $2,3$ for Rindler generators. The  deformed Poincar\'e Casimir can be now used to obtain a deformed Rindler wave operator
\begin{widetext}
\be
\label{acc.disp.rel.}
\mathcal{C^{\mathcal{F}}}=\frac{e^{-2a \xi}}{(1-\ell P^{\mathcal{F}}_\eta)^2}\left[ -(P^{\mathcal{F}}_\eta)^2+(P^{\mathcal{F}}_\xi)^2+(P^\mathcal{F}_\perp)^2(P^{\mathcal{F}}_\eta\ell+(1-\ell P^{\mathcal{F}}_\eta)e^{a \xi})^2\right]\,. 
\ee
\end{widetext}

As shown in Section 2, in order to derive a formula for the Doppler shift, one first derives the four-momentum (four-velocity) of an accelerated observer by anti-boosting by $\alpha \tau$ the rest four-momentum $(m,0,0,0)$. This turns out to be ``parallel" to the Rindler time-translation Killing vector and the proportionality factor is responsible for the Doppler shift between accelerated trajectories with different accelerations. In the case of translation generators belonging to a deformed algebra we deal with non-linear redefinitions of ordinary Lie algebra generators and thus the vector space structure and with it the notion of parallelism between vectors are no longer available. 

As stated at in Section 2 our approach will be to {\it define} the deformed Rindler frequency shift simply as a dilation transformation albeit, in this case, deformed according to the non-linear structure of the twisted Weyl-Poincar\'e algebra\footnote{Notice that this is indeed what one would obtain by mapping the undeformed Rindler redshift \eqref{RshiftO} under the twist map \eqref{twistedPWG}.}.  The resulting deformed Doppler shift formula is given by an inverse dilation 
\be
\label{def.red.fact.}
\omega_\xi=\frac{\omega}{\ell\omega+(1-\ell\omega)e^{a \xi}}\,.
\ee
which trivially reduces to the usual shift \eqref{RshiftO} in the limit of vanishing $\ell$. The new feature due to the deformation of the Weyl-Poincar\'e algebra is that in the limit of an infinite contraction $\xi \rightarrow -\infty$, the ``Rindler horizon limit", the frequency approaches the maximal Planckian value of $1/\ell$. Thus we see that the putative Plank-scale effects encoded in the symmetry deformation realize the common expectation of ``smoothing" out the Rindler horizon by providing a fundamental frequency cut-off scale. It is natural to expect that this finite behaviour will reflect on the state counting for a Rindler field, taming the divergence described in Section 3. This is the analysis that we carry out in the next Section.

\section{Mode counting for deformed fields}
In Section 3 we showed that in order to perform state counting in momentum space we need a mass-shell condition and an integration measure on momentum space. While we have derived deformed mass-shell conditions both for deformed Minkowski and Rindler translation generators we have not discussed so far whether the momentum space metric should be affected by the deformation.
Unlike the case of the $\kappa$-Poincar\'e algebra, where it is well known that momentum space is ``half" of de Sitter space, the geometry of the momentum space associated to the twisted Weyl-Poincar\'e algebra does not have a clear characterization. A first natural guess for a momentum space measure would be just to adopt the usual flat Lebesgue measure\footnote{In this and the following sections, for obvious notational convenience, we will drop the superscript $\mathcal{F}$ from deformed Minkowski and Rindler momenta.} $d\mu(p) = d^4 p$. It turns out that such measures {\it is not} invariant under the deformed boosts \eqref{energy.boost.}-\eqref{moment2.boost.}. This feature has been recently noted in \cite{Gubitosi:2015jcj} elaborating on the original Magueijo-Smolin DSR model. It turns out that, as one could have easily guessed, a covariant measure can be obtained from a  transformation of the ordinary flat measure into the following deformed measure via the twist map \eqref{twistedPWG}:
\be
\label{inv.curv.meas.}
d\mu(p)=\frac{d^4 p}{(1-\ell p_0)^5}.
\ee
In what follows, for completeness, we will consider both measures since, as we will see, the Lorentz breaking measure $d\mu(p) = d^4 p$ will unexpectedly give the most interesting results. 

As a warm up let us start from the counting of states for a field with a deformed momentum space associated to the Weyl-Poincar\'e algebra. As extensively reviewed in Section 3 we will consider the following density of states
\be
n(E)=\frac{V}{(2\pi)^3}\int_E d\mu(p) \, 2 p_0\, \delta(\mathcal{C})\, \theta(p_0)\,,
\ee
with the mass-shell $\mathcal{C}$ given by \eqref{def.cas.}, for both choices of momentum space measure $d\mu(p)$. In the case of the Lorentz breaking flat measure $d\mu(p) = d^4 p$, one obtains
\be\label{dens.states.inert.flat.}
n(E)_{\mathrm{LIV}} = \frac{2V}{(2\pi)^2}\left[ \frac{E^3}{3}-\frac{\ell E^4}{2}+\frac{\ell^2 E^5}{5} \right]\,,
\ee
while for the measure \eqref{inv.curv.meas.} we get
\be\label{dens.states.inert.curv.}
n(E)_{\mathrm{C}}=\frac{V}{(2\pi)^2}\frac{1}{\ell^3}\left[ \frac{\ell E(3\ell E-2)}{(1-\ell E)^2}-2\log(1-\ell E) \right].
\ee
These expressions can be used to derive all relevant deformed thermodynamical quantities of interest. What is interesting to check is whether the number of states remains finite in the limit of Planckian (maximal) energy $1/\ell$ or not. As it can be easily checked, in the case of a Lorentz violating measure, 
\be
\lim_{E\to1/\ell} n(E)_{\mathrm{LIV}} =\frac{V}{(2\pi)^2}\frac{1}{15\ell^3}\,,
\ee
i.e we have a finite number of states all the way up to the Planck scale, while in the fully covariant picture,
\be
\lim_{E\to1/\ell} n(E)_{\mathrm{C}} = \infty
\ee
showing that, despite the deformation and the maximal value of the energy, the number of states available to the system is still infinite. As we will see below this divergence will haunt us also in the deformed Rindler case.

As a next step we evaluate the entropy associated to both density of states. This can be done considering the logarithm of the partition function
\be
\log Q=\beta\int_0^{1/\ell} dE \, \frac{n(E)}{e^{\frac{\beta E}{1-\ell E}}-1} \, ,
\ee
with the deformed Bose-Einstein distribution $1/(e^{\frac{\beta E}{1-\ell E}}-1)$ the natural choice determined by the twist map \eqref{twistedPWG}. In the Lorentz breaking case, from \eqref{dens.states.inert.flat.} we find the deformed entropy 
\begin{align}
S_{\mathrm{LIV}}
& =-\beta^2\frac{\partial}{\partial\beta}\frac{\log Q_{\mathrm{LIV}}}{\beta}= \nonumber \\
\label{entropylivinert}
& =\frac{V}{(2\pi)^2}\frac{\beta^2}{15}\int_0^\infty dx \, \frac{x^4(10+5\ell x+\ell^2 x^2)}{(1+\ell x)^5}\frac{e^{\beta x}}{(e^{\beta x}-1)^2} \,.
\end{align}
where in the second line we performed the change of integration variable $x=E/(1-\ell E)$. Since in the above equation the integral converges, the entropy $S_{\mathrm{LIV}}$ is finite. From \eqref{dens.states.inert.curv.}, we can similarly write the entropy associated to the covariant measure
\be \label{entropycovinert}
S_\mathrm{C}=\frac{V}{(2\pi)^2}\frac{\beta^2}{\ell^3}\int_0^\infty dx \, \frac{ x\left[ \ell x(\ell x-2)+2\log(1+\ell x) \right]e^{\beta x}}{(e^{\beta x}-1)^2} \,.
\ee
which is again finite. Even though we are not able to explicitly solve integrals \eqref{entropycovinert} and \eqref{entropylivinert}, we can expand them in series for $\ell \rightarrow 0$ obtaining 
\begin{align}
S_{\mathrm{LIV}}=\frac{2\pi^2}{45\beta^3}-\ell\frac{10\zeta(5)}{\pi^2\beta^4}+\mathcal{O}(\ell^2) \,, \\
S_{\mathrm{C}}=\frac{2\pi^2}{45\beta^3}-\ell\frac{15\zeta(5)}{\pi^2\beta^4}+\mathcal{O}(\ell^2) \,,
\end{align}  
where $\zeta(n)$ is the Riemann Zeta function. We see that the zeroth order term in both cases reproduces the known result for a gas of massless scalar particles while the firsts corrections of order $\ell$ in the two cases differ for a numerical factor.

We now move to the Rindler case. To evaluate the density of states we start from the following generalization of \eqref{nRindU}
\be
n(E) = \frac{V_\perp}{(2\pi)^3}\int_{\mathbb{R}} \, d\xi \int \, dp_\eta dp_\xi dp^2_\perp \, 2p_\eta e^{-2a\xi} \,  \delta(\mathcal{C})\theta(p_\eta)\,.
\ee
In the deformed case we will consider the integral over $\xi$ as an integral over all possible values of the dilation parameter. In the integral over momentum space we will have to introduce the deformed Rindler mass-shell \eqref{acc.disp.rel.}, a possibly non-trivial measure over deformed Rindler momenta and replace $e^{-2a\xi}$ with the square of the deformed Doppler shift factor 
\be
\frac{1}{\ell\omega+(1-\ell\omega)e^{a \xi}}\,,
\ee
obtained from \eqref{def.red.fact.}. The resulting integral is \\
\begin{widetext}
\begin{align}
n(E) \label{nerindef}
& =\frac{V_\perp}{(2\pi)^3}\int_{\mathbb{R}} \, d\xi \int \, d\mu(p_\eta,p_\xi,p_\perp) \, \frac{2p_\eta}{(\ell p_\eta+(1-\ell p_\eta)e^{a\xi})^2} \, \delta(\mathcal{C})\theta(p_\eta)=\nonumber \\
& =\frac{V_\perp}{(2\pi)^3}\int_{\mathbb{R}} \, d\xi \int \, d\mu(p_\eta,p_\xi,p_\perp) \, \frac{p_\eta (1-\ell p_\eta)^2 e^{2a\xi}}{(\ell p_\eta+(1-\ell p_\eta)e^{a\xi})(p_\eta e^{a\xi}+\ell p_\xi^2 (1-e^{a\xi}))} \, \delta(p_\eta-\omega_p)\theta(p_\eta),
\end{align}
\end{widetext}
where $\omega_p$ is the positive solution of the Rindler mass-shell \eqref{acc.disp.rel.}. Carrying out the calculation using the covariant measure we obtain the following {\it diverging} density of states\\
\be
n(E)_{\mathrm{C}} =\frac{V_\perp}{(2\pi)^3}4\pi \int_0^E du \, u^2(1-\ell u)^2 \int_{-\infty}^\infty d\xi \, a e^{-2 a\xi}.
\ee
To proceed we need to introduce a ``brick wall" regulator at $\xi = \xi_{min}$ and, carrying out the energy integral, we obtain
\be\label{dens.state.acc.cov.}
n(E)_{\mathrm{C}} = \frac{V_\perp}{(2\pi)^2}\frac{e^{-2a\xi_{min}}}{a} \left[ \frac{E^3}{3}-\frac{\ell E^4}{2}+\frac{\ell^2E^5}{5} \right] \,.
\ee
Quite interestingly, evaluating the integral in \eqref{nerindef} using the Lorentz breaking measure we obtain the following {\it finite} density of states
\be\label{dens.state.acc.flat.}
n(E)_{\mathrm{LIV}} =-\frac{V_\perp}{(2\pi)^2}\frac{1}{6a}\frac{1}{\ell^3}\log(1-\ell E),
\ee
which diverges in the limit of vanishing $\ell$ thus reproducing the standard result. We thus come to the conclusion that, despite the finite blueshift introduced by the symmetry deformation, a finite density of states can be only obtained, in the particular model considered, by introducing features which break Lorentz invariance.

It is instructive to evaluate the entropy associated to the densities of states we just derived for a deformed Rindler field. For the covariant measure, from \eqref{dens.state.acc.cov.}, we obtain 
\be
S_\mathrm{C}=\frac{V_\perp \beta^2 e^{-2a\xi_{min}}}{60 \pi^2 a}\int_0^\infty dx\, \frac{x^4(10+5\ell x+\ell^2 x^2)}{(1+\ell x)^5}\frac{e^{\beta x}}{(e^{\beta x}-1)^2} \,,
\ee
where, again we set the integration variable to $x=E/(1-\ell E)$, while for the Lorentz breaking measure, from \eqref{dens.state.acc.flat.} we get
\be
S_\mathrm{LIV}=\frac{V_\perp}{(2\pi)^2}\frac{1}{6a}\frac{\beta^2}{\ell^3}I \,,
\ee
with
\be
I=\int_0^\infty dx \, \frac{x\log(1+\ell x)e^{\beta x}}{(e^{\beta x}-1)^2}
\ee
a finite integral. Thus, as expected, the entropy of the deformed Rindler field in the presence of the Lorentz breaking measure converges. Let us write down the leading terms in these two entropies, and check whether they reproduce the area-law when are evaluated at the Unruh temperature $T_U=a/2\pi$. We have
\begin{align}
S_{\mathrm{C}}|_{\beta=\frac{1}{T_U}}
& =\frac{V_\perp e^{-2a\xi_{\min}}}{a}\left[\frac{\pi^2}{45\beta^3}-\ell\frac{5\zeta(5)}{\pi^2\beta^4}+\mathcal{O}(\ell^2)\right]_{\beta=\frac{1}{T_U}}= \notag \\
& =\frac{V_\perp a^2 e^{-2a\xi_{\min}}}{360 \pi}-\ell \frac{V_\perp a^3 5 \zeta(5) e^{-2a\xi_{\min}}}{16\pi^6}+\mathcal{O}(\ell^2) \,, \\
S_\mathrm{LIV}|_{\beta=\frac{1}{T_U}}
& =\frac{V_\perp}{a}\left[\frac{1}{72\beta\ell^2}-\frac{\zeta(3)}{8\pi^2\beta^2\ell}+\frac{\pi^2}{270\beta^3}+\mathcal{O}(\ell) \right]_{\beta=\frac{1}{T_U}}= \notag \\
\label{entropylivrindler}
& =\frac{V_\perp}{144\pi\ell^2}-\frac{V_\perp a}{\ell}\frac{\zeta(3)}{32\pi^4}+\frac{V_\perp a^2}{2160\pi}+\mathcal{O}(\ell) \,.
\end{align}
In both cases we reproduce the area law $S \propto A \equiv V_\perp$, as in the Bekenstein-Hawking formula. For $S_{\mathrm{C}}$ the zeroth order contribution in $\ell$ depends on the brick-wall length scale $\epsilon\sim e^{a\xi_{\min}}/a$ in analogy with the result of \cite{Susskind-Uglum:1994}. It is interesting to note in the zeroth order term in $S_\mathrm{LIV}$ the deformation scale $\ell$ plays the role of natural regulator and for $\ell_P=6\sqrt{\pi}\ell$ reproduces the Bekenstein-Hawking relation \eqref{BES}.

\section{Discussion} 

In this work we provided a characterization of the kinematics of accelerated observers within the framework of deformations of relativistic symmetries. These deformations have been widely studied in the literature as modifications of ordinary relativistic kinematics encoding putative Planck scale effects possibly emerging in a ``flat-space-time limit" of quantum gravity. We focused in particular on modifications of the frequency shift associated with Rindler space translation generator and the density of states of a field near the accelerated horizon. 

We started by recasting the construction of Rindler space and time translation generators in terms of the action of the Weyl-Poincar\'e group and evidencing the relationship between space-time dilation transformations and the notion of accelerated horizon seen as the locus reached with an infinite inverse dilation.  We then considered a specific example of deformation of the Weyl-Poincar\'e algebra constructed via a twist procedure, recently considered in the literature. A first result was the derivation of finite boosts and dilation transformations associated to the deformed algebra which evidenced that energy and momentum reach a finite, Planckian, value in the limit of boost and dilation parameter going to infinity. A similar behaviour for boosts was first explored in the context of DSR models in which the Planck energy is introduced as an observer independent energy scale \cite{AmelinoCamelia:2000ge, AmelinoCamelia:2000mn}. Quite interestingly the deformed boost transformations which we derived coincide with those proposed in one of the early DSR works \cite{Magueijo:2001cr}. Our results also exposed a similar behaviour for dilations which reflects in a {\it finite} blueshift for the frequency of a field mode in the limit of infinite inverse dilation, i.e. at the Rindler horizon.

Having laid down the basics of deformed Rindler kinematics we explored the behaviour of a massless field close to the Rindler horizon. In particular we focused on the density of states, used to derive the area dependent entropy in the brick-wall model, which in ordinary Rindler space diverges due to the infinite blueshift at the horizon.
We found that, despite the finite frequency shift at the horizon, adopting an integration measure in momentum space covariant under deformed boosts, one obtains a diverging density of states. Adopting instead an ordinary momentum space measure, and thus violating Lorentz invariance, produces a finite density of states. Thus in the particular deformation of the Weyl-Poincar\'e we considered the volume of phase space described by the covariant measure leads to a divergent density of states (and associated thermodynamic quantities). This, of course, does not exclude the possibility that there might be other types of deformation of the Weyl-PoincareÕ algebra which might  lead to a finite density of states at the Rindler horizon without resorting to the introduction of features which break Lorentz symmetry. Exploring such possibility would indeed be an interesting direction for future work.

Finally, it is interesting to note that the analysis we presented suggests that the commonly accepted view that quantum gravity should lead to a finite entropy density associated to any local Rindler horizon, supports the case for a departure from the ordinary description of relativistic symmetries at the Planck scale, be it in the form of a breaking or deformation of Lorentz symmetry.  Hopefully this could provide some valuable insight on the current puzzles concerning black hole quantum radiance.\\ 



\end{document}